\documentclass[bibyear]{aa}  

\usepackage{graphicx}
\usepackage{txfonts}
\begin{document}

\title{Hot coronal loops associated with umbral brightenings}
\author{C. E. Alissandrakis and S. Patsourakos}

\institute{Department of Physics, University of Ioannina, GR-45110 Ioannina, 
Greece\\
\email{calissan@cc.uoi.gr, spatsour@cc.uoi.gr}
}

\date{Received Apr 26, 2013; accepted Jun 21, 2013}

 
  \abstract
   {}
   {
We aim to investigate the association of umbral brightenings with coronal structures.}
   {We analyzed AIA/SDO high-cadence images in all bands, HMI/SDO data,  soft X-ray 
images from SXI/GOES-15, and H$\alpha$ images from the GONG network.} 
   {We detected umbral brightenings that were visible in all AIA bands as well as 
in H$\alpha$. Moreover, we identified hot coronal loops that connected the 
brightenings with nearby regions of opposite magnetic polarity. These loops were 
initially visible in the 94\,\AA\ band, subsequently in the 335\,\AA\ band, and 
in one case in the 211\,\AA\ band. A differential emission measure analysis revealed 
plasma with an average temperature of about $6.5\times10^6$\,K. 
This behavior suggests cooling of impulsively heated loops.} 
{}


   \maketitle
%

\section{Introduction}
Sunspots are the site of many dynamic phenomena, such as umbral oscillations, 
running umbral/penumbral waves, and umbral flashes (see review by Solanki 
\cite{2003A&ARv..11..153S}). Beckers \& Tallant (\cite{1969SoPh....7..351B}) used 
the term {\it umbral flashes} to describe small (average diameter of 2200~km), 
short-lived ($<120$~s) bright structures that they observed in the K line. They 
noticed a tendency of these flashes to repeat every 145\,s, which led many subsequent authors 
to identify them with the {\it umbral oscillations} discovered later by Bhatnagar 
\& Tanaka (\cite{1972SoPh...24...87B}), Beckers and Schultz (\cite{1972SoPh...27...61B}) 
and Giovanelli (\cite{1972SoPh...27...71G}); however, the time profiles of umbral 
flashes presented by Beckers \& Tallant (\cite{1969SoPh....7..351B}) did not have 
a sinusoidal form, but 
showed a fast rise followed by a slower decay with a characteristic time of 
$\sim50$~sec. 

Umbral flashes are difficult to observe in other chromospheric lines in the 
visible part of the spectrum. One case in H$\alpha$ was reported by Alissandrakis 
et al. (\cite{1992SoPh..138...93A}). These  authors were also the first to report 
on waves propagating from inside the umbra outward  to become 
penumbral waves (Zirin \& Stein \cite{1972ApJ...178L..85Z}; Giovanelli 
\cite{1972SoPh...27...71G}). The propagating nature of umbral waves 
was subsequently verified by Alissandrakis et al. (\cite{1998ASPC..155...49A}) and 
Tsiropoula et al. (\cite{2000A&A...355..375T}).

Another type of umbral brightenings, termed {\it umbral flares}, was reported by 
Tang (\cite{1978SoPh...60..119T}). They appear in H$\alpha$ as bright patches 
confined inside the umbra, with the other footpoint located in the nearby plage;
they last for 25-45 min and are sometimes accompanied by type III bursts.

An important aspect of sunspot-associated dynamic phenomena is their extension in 
the upper layers of the solar atmosphere. Indeed, Gelfreikh et al. (\cite{1999SoPh..185..177G}) 
detected sunspot oscillations at 1.7 cm with the Nobeyama radioheliograph, while
Nindos et al. (\cite{2002A&A...386..658N}) resolved them spatially using the VLA. 
The availability of continuous high-cadence, high-resolution observations of the 
entire sun from instruments onboard the Solar Dynamics Observatory (SDO) gives an 
excellent opportunity for detailed studies of sunspot-associated phenomena in the 
transition region and the corona ({\it e.g.\/} Reznikova et al. \cite{2012ApJ...746..119R}). 

In this article we used AIA/SDO data to investigate the extension of umbral 
brightenings into the transition region and the low corona. We describe the 
observations in Section 2 and discuss the results in Section 3.

\section{Observations and results}
We examined SDO/AIA images of a number of isolated, well-developed symmetric sunspots 
with the help of the {\it Helioviewer} site for the occurrence of brightenings 
above their umbra. We found several cases, best visible in the 1600\,\AA\ band, and 
selected two for more detailed study (Table \ref{table:1}); for this study we used 
cutouts of the original high-cadence images.

\begin{table}[h]
\caption{List of Observations}             
\label{table:1}      
\centering                          
\begin{tabular}{l c c c}        
\hline\hline                 
Date        &Region& Location    & Peak UT\\    
\hline                        
2013 Jan 19 & 1658 & W15.7 S12.0 & 19:14 \\
2012 Sep 30 & 1579 & E01.8 S10.0 & 07:53 \\  
\hline                                   
\end{tabular}
\end{table}

\subsection{Event of January 19, 2013}
\begin{figure*}
\centering
\includegraphics[width=\textwidth]{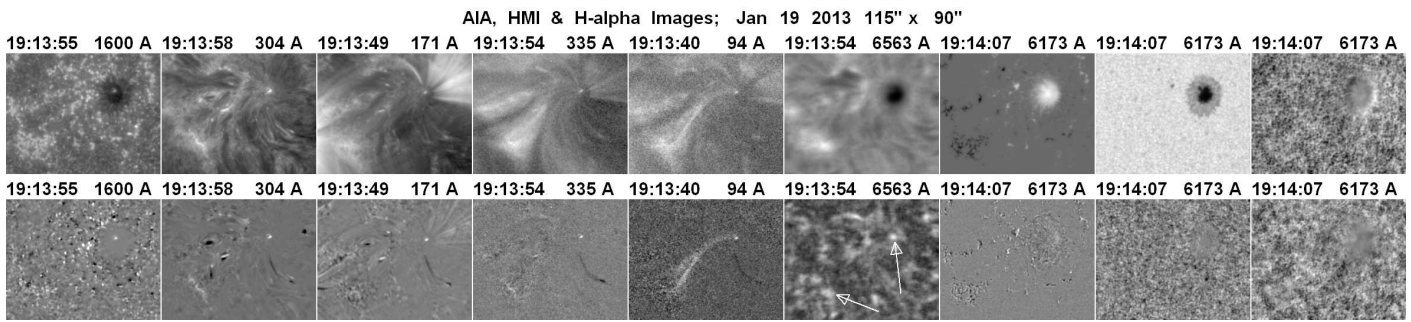}
\caption{Umbral brightening of January 19, 2013 in selected AIA bands and 
H$\alpha$, together with HMI images of the longitudinal magnetic field, continuum 
intensity, and line-of-sight velocity. The bottom row shows difference images. 
The arrows in the H$\alpha$ difference image point to the umbral brightening 
and the associated plage footpoint. The images are orientated in the solar 
E-W/N-S direction. A time sequence of images is shown in movie 1 in the electronic
version of the journal. Each frame of the movie shows images 
in the 1600 and 304\,\AA\ bands (top row) as well as in the  94 and 335\,\AA\ 
AIA bands (bottom row); the first frame shows direct images in all four 
bands, the others show difference images in all bands except 1600\,\AA.}
\label{Fig1}%
\end{figure*}

A well-observed case is the event of January 19, 2013. Figure \ref{Fig1} shows a 
set of images in selected AIA wavelength bands together with HMI images of 
continuum intensity, longitudinal magnetic field, and line-of-sight velocity, as 
well as an H$\alpha$ image from the GONG network, during the peak of the 
brightening. The bottom row shows images from which the intensity before the event 
was subtracted for better visibility of the brightening.

The brightening, located 3\arcsec\ N of the spot center, 
is visible in all AIA bands; it is barely visible in the lower-resolution 
H$\alpha$ difference image and there is no trace of it in the continuum, $B_\ell$ 
and $v_\ell$ images, not even in the image of the line depth of the Fe{\sc i} 
6173\,\AA\ line used by HMI (not shown here). Thus the lower extent of the 
brightening is somewhere between the formation height of the core of the Fe{\sc i} 
line ($\sim302$ km, see Norton et al. \cite{2006SoPh..239...69N}) and of the 
1700\,\AA\ band ($\sim360$ km, see Fossum \& Carlsson \cite{2005ApJ...625..556F}). 
We found no indication of associated type III burst activity in the CALLISTO and 
WIND/WAVES data bases. The peak intensity of the brightening relative to the one 
before the event is given in Table \ref{table:2}; the brightening is strongest in 
the 1600\,\AA\ band where its intensity is higher than that of the quiet Sun and 
similar to the average intensity of plage regions; the brightening is weakest in the 
1700\,\AA\ band.

\begin{table}[h]
\caption{Peak intensity of brightenings relative to the local background}             
\label{table:2}              
\centering                   
\begin{tabular}{r c c c}     
\hline\hline                 
Band        &2013  &2012      &2012       \\    
            &Jan 19&Sep 30, E &Sep 30, W  \\    
\hline                                           
1700        & 1.86 &   --     & 1.53      \\
1600        & 4.70 & 1.73     & 3.25      \\
 304        & 3.74 & 3.00     & 3.08      \\
 171        & 3.12 & 1.25     & 2.94      \\
 211        & 2.66 & 1.90     & 1.48      \\
 193        & 2.53 & 1.62     & 2.30      \\
 131        & 3.26 & 2.50     & 3.08      \\
 335        & 2.81 & 1.78     & 2.25      \\
  94        & 3.00 & 2.26     & 1.80      \\
\hline                                   
\end{tabular}
\end{table}

\begin{figure}[h]
\centering
\includegraphics[width=8.95cm]{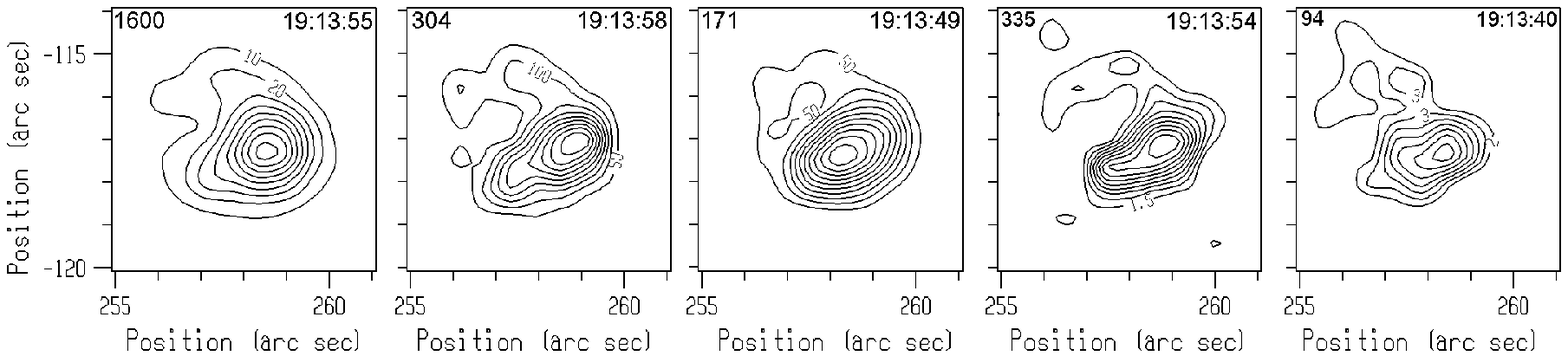}\\
\vspace{.1cm}
\hspace{0.7cm}\includegraphics[width=8.3cm]{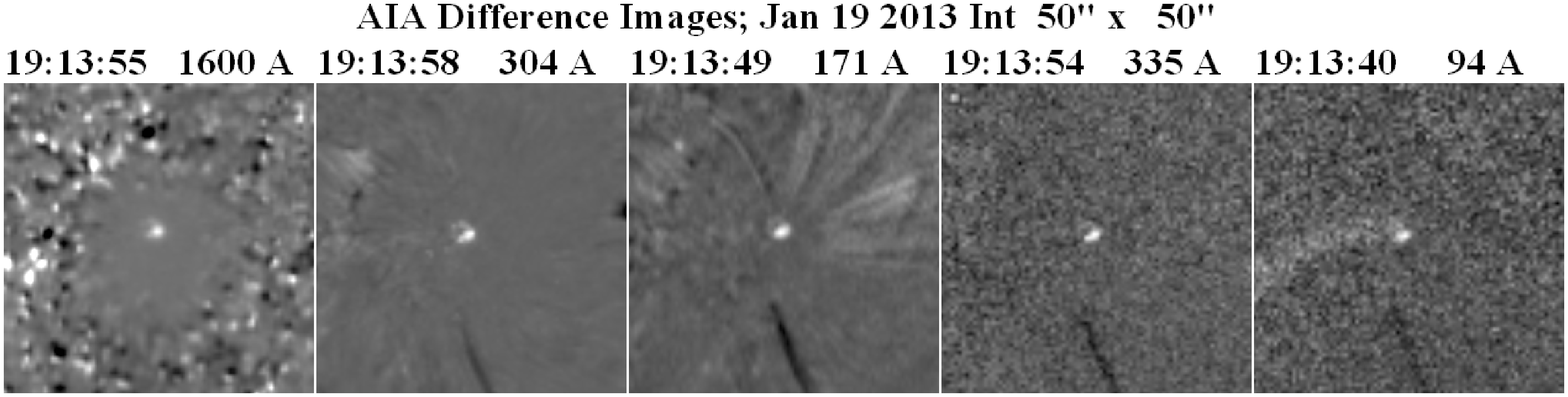}
\caption{Contour plots and difference images of the brightening at the time of 
 its maximum.}
\label{structure}
\end{figure}

At the time of its peak, the brightening had a rather simple structure (Figure 
\ref{structure}), with small differences among the AIA spectral bands. It was 
roughly elliptical in shape, with a FWHM of 2.4\arcsec\ by 1.4\arcsec\ in the 
171\,\AA\ image, with an arch-like extension NE; the structure was more compact 
in the 304, 335, and 94 \AA\ images, with the peak displaced slightly to the NW 
and some indication of a second peak.

The most interesting aspect of the brightening is that in the 94\,\AA\ band image, 
with a characteristic temperature of $\log T\sim 6.8$ (Lemen et al. 
\cite{2012SoPh..275...17L}), it is clearly associated 
with a coronal loop (Figure \ref{Fig1}), best seen in the subtracted image.
The loop ends  in a region of opposite polarity at a projected distance 
of 53\arcsec, which is bright in H$\alpha$, and  in the 304\,\AA\ and some 
other AIA bands. This opposite footpoint is more extended than the brightening, 
and the cross-section of the loop along its length varies accordingly. This is 
consistent with the expected variation of the magnetic field strength from the 
sunspot to the plage. 

\begin{figure}
\centering
\includegraphics[width=\hsize]{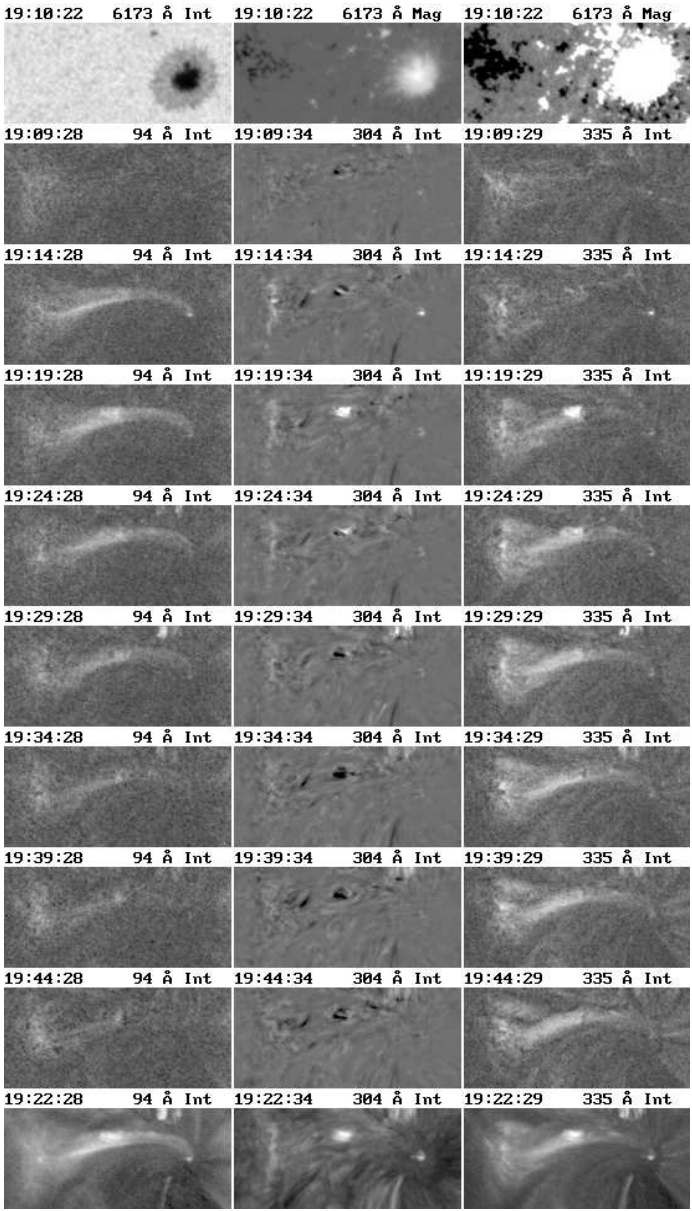}
\caption{Time sequence of difference images in the AIA 94, 304, and 
335\,\AA\ bands over a field of view of 100\arcsec\ by 45\arcsec. The last row 
shows images of the intensity rms during the sequence (19:00 to 19:45 UT), while 
the first row shows HMI images of continuum intensity and longitudinal magnetic 
field, saturated at $\pm100$\,G in the right column. The 335 and 94\,\AA\ images are 
one minute averages to reduce noise. The small bright structure near the middle 
of the loop is an unrelated phenomenon. The images have been rotated by 39\degr\ 
with respect to the solar E-W direction.}

\label{seq}
\end{figure}

The loop developed together with the brightening and  was not detected in any 
other spectral band up to the maximum of the brightening. After that, 
the loop became progressively visible in the 335\,\AA\ band (characteristic 
temperature of $\log T\sim6.4$), while it faded away in the 94\,\AA\ band (Figure 
\ref{seq}). Note that the loop persisted for several minutes after the brightening 
had faded away. Also note that the overall geometry of the loop and its footpoints 
is very well depicted in the images of the rms variation of the intensity (last 
row in Figure \ref{seq}). 

No observations from Hinode were available at the time of the event. Soft X-ray 
images from SXI on GOES-15 show the loop in all channels, starting around 09:10~UT 
with a peak near the peak of the umbral brightening.

\begin{figure}[h]
\centering
\hspace{0.3cm}%
\includegraphics[width=2.5cm]{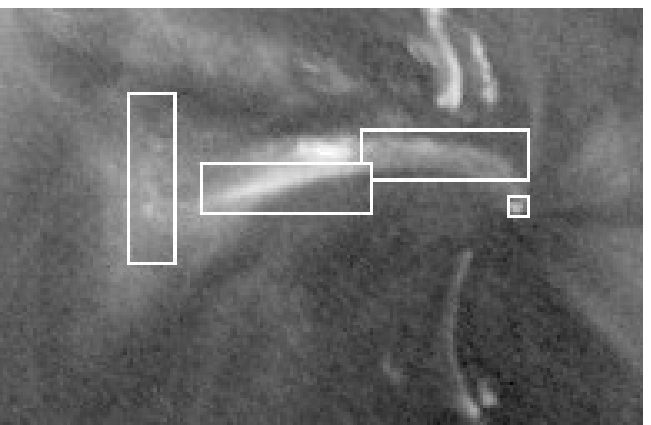}\hspace{.35cm}%
\includegraphics[width=2.5cm]{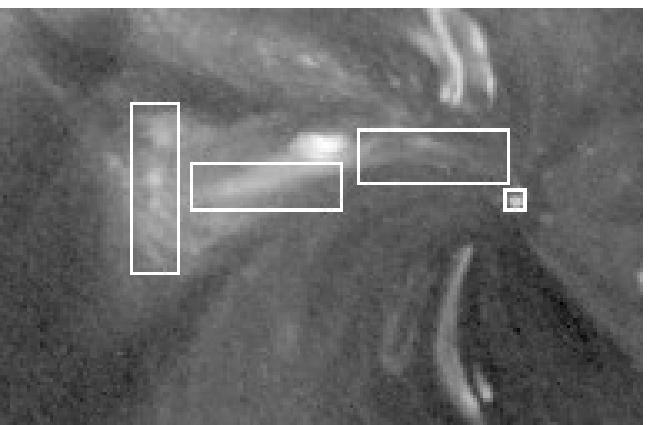}\hspace{.44cm}%
\includegraphics[width=2.5cm]{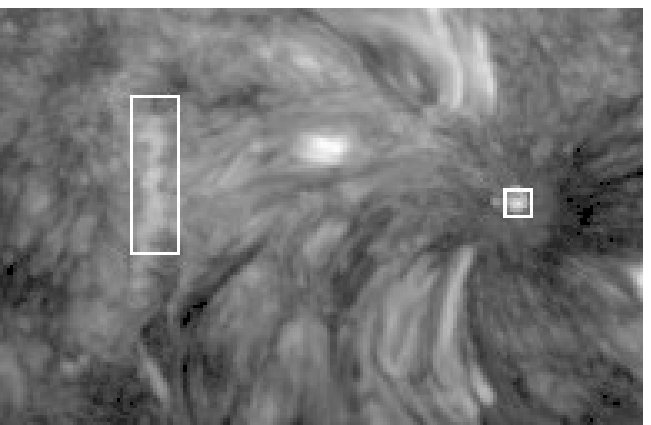}\\
\vspace{.2cm}
\includegraphics[width=\hsize]{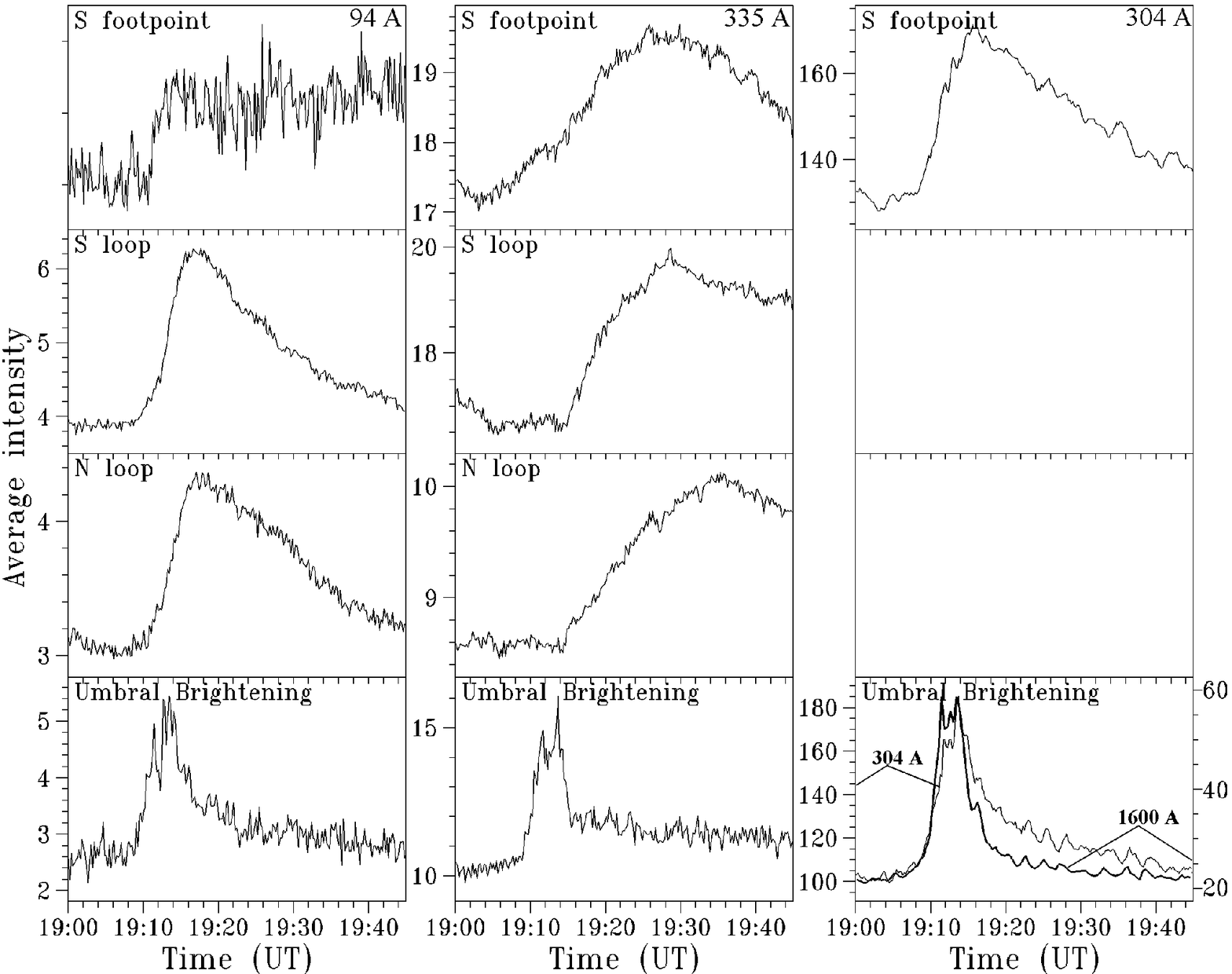}
\caption{Time profiles of the brightening, the southern footpoint and two portions of 
the loop. The boxes in the top row images show the corresponding regions  in rms 
images (from left to right: S footpoint, S loop, N loop, umbral brightening).}
\label{timprof}
\end{figure}

Figure \ref{timprof} shows the time profiles of the brightening, of two portions 
of the loop and of its southern footpoint. The time profiles of 
the brightening and the S footpoint during the decay phase in 94 and 335 \AA\ 
are probably contaminated by the loop emission. The brightening had two main 
peaks, a fast rise of a few minutes and a much longer decay. The duration of the 
main phase was shortest in the 1600\,\AA\ band ($\sim10$~min). The S footpoint 
appeared almost simultaneously with the brightening at 304\,\AA, with a delay of 
$\sim1$ min in the rising phase and $\sim3$ min in the peak. 

In the 94\,\AA\ band the S and the N portions of the loop showed practically 
identical light curves, both delayed by $\sim6$ min with respect to the peak of 
the brightening. In 335\,\AA\ the S footpoint appeared first, $\sim12$~min 
after the peak of the brightening, followed by the S part of the loop after 
$\sim3$~min and the N part of the loop $\sim7$ min later. Compared with the 
94\,\AA\ band, the N part of the loop in 335\,\AA\ was delayed by $\sim18$ min
and the S part by $\sim12$~min.

\begin{figure}
\centering
\includegraphics[width=\hsize]{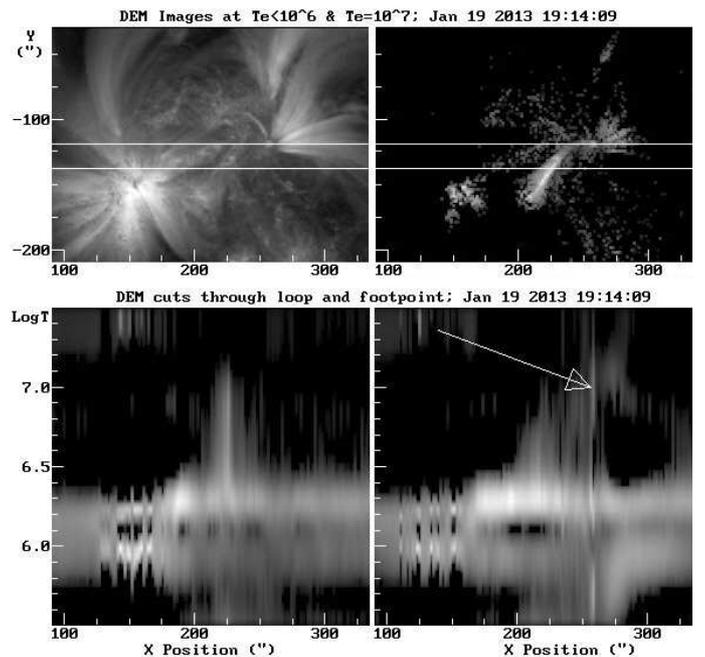}
\caption{Differential emission measure maps for $T<10^6$\,K and $T=10^7$\,K (top).
The bottom row shows the DEM as a function of position and temperature along the 
lines marked in the images at the top. The arrow marks the position of the 
brightening.}
\label{DEM13}
\end{figure}

To quantify the temperature structure of the loop we performed a 
differential emission measure (DEM) analysis based on the coronal AIA channels, 
using the method described in \cite{2013ApJ...771....2P}. 
Our results are presented in Figure \ref{DEM13}, 
which shows DEM images of the region for temperatures below 10$^6$\,K and at 
10$^7$\,K; in the first image the loop is invisible, while in the other it 
stands out clearly above the surrounding plasma. In the DEM images as a function 
of position and temperature shown in the bottom row of the figure, the bulk of 
the background plasma is in the temperature range $5.75 < \log T < 6.4$. In the 
lower left panel, where the DEM is displayed along a line crossing the loop at 
its middle, the loop appears at $x\sim225$ as an excursion of the DEM to high 
temperatures; the DEM peaks at $\log T\sim 6.82$ ($T \sim 6.5 \times 10^6$\,K). 
The lower right panel of Figure \ref{DEM13} shows the DEM along a line crossing 
the umbral brightening at $x=258$; here the DEM also extends to high temperatures, 
with a maximum at $\sim \log T\sim 6.96$.

\begin{figure*}[t]
\centering
\includegraphics[width=12cm]{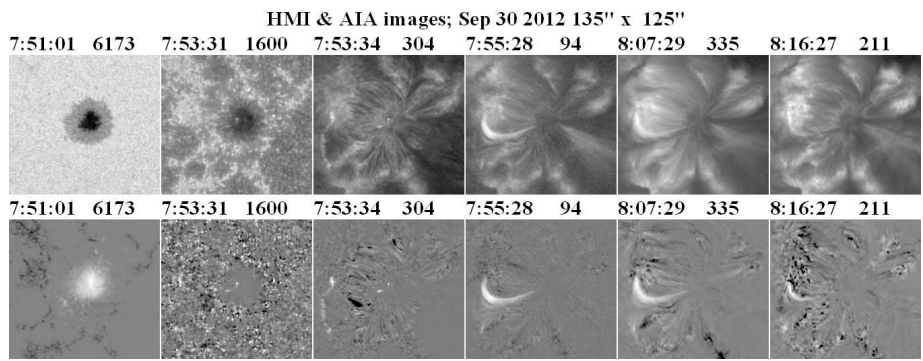}
\caption{Umbral brightening of September 30, 2012 in selected AIA bands. 1600 
and 304 \AA\ images are at the peak of the brightening, 94, 335, and 211 \AA\ 
images are at the peak of the associated coronal loop. 94 and 335 \AA\ images are 
1-min averages to reduce noise. The bottom row shows difference images. The first 
column shows HMI images of continuum intensity and longitudinal magnetic field. 
The time evolution of this loop and its footpoins is shown in movie 2
of the electronic version of the journal. Each frame of the movie shows 
difference images in the 1600 and 304\,\AA\ bands (top row) as well 
as in the  94 and 335\,\AA\ AIA bands (bottom row).}
\label{sep30}
\end{figure*}

\subsection{Event of September 30, 2012}
A similar event was observed on September 30, 2012 in active region 1759 (Figure 
\ref{sep30}). This brightening consisted of two patches, 10\arcsec\ apart, 
connected by a faint emission bridge that gave them a sickle-like appearance. The 
E patch started near the umbra-penumbra boundary and extended  to the middle of 
the penumbra, the W patch was entirely within the umbra, $\sim$5\arcsec\ away 
from its center; their sizes were roughly 4 by 1.5\arcsec\ and  5 by 2\arcsec\ 
respectively, but both contained up to four smaller components, down to 1\arcsec\ 
size. A very weak type III burst was discernible during the rise phase of the 
W patch, at 07:51:09 UT in the San Vito dynamic spectra below 60\,MHz and in the 
WIND/WAVES RAD2 spectra (07:51 UT at 13.8\,MHz and 07:52 UT at 2\,MHz); lacking 
imaging data, we cannot associate them with certainty with the event.

The coronal loop started near the umbra-penumbra boundary, at the west end of the 
E patch and extended over 44\arcsec\ to the adjoining opposite polarity plage. The plage 
footpoint was visible not only in the 304\,\AA\ band, but also in the 
1600\,\AA\ band (Figure \ref{sep30}); it was also visible in the GONG difference 
H$\alpha$ image, where the brightening itself was not. The time evolution of this 
loop is similar to that of the event of January 19, 2013: it was best visible in 
the 94\,\AA\ band, $\sim$2 min after the maximum of the umbral brightening and in 
the 335\,\AA\ band $\sim$12 min later. Here we also had emission in the 211\,\AA\ 
band (characteristic temperature of $\log T\sim6.3$), which peaked $\sim$23 min 
after the brightening. Soft X-ray images from GOES SXI showed a picture similar 
to that of the January 19 2013 event: the hot loop was visible in all SXI 
channels and appeared at about the same time as the 94\,\AA\ loop. 

The light curve of the brightening itself had a fluctuating character, with the bulk of the 
emission lasting for $\sim$8 min in the 1600\,\AA\ band. The peak intensities of 
the E and W components of the brightening relative to the local background are 
given in Table \ref{table:2}.

\begin{figure}
\centering
\includegraphics[width=\hsize]{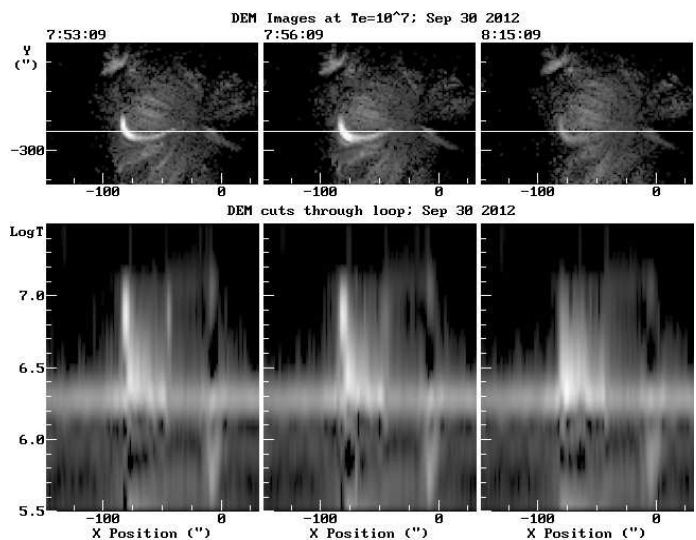}
\caption{Differential emission measure maps for $T=10^7$\,K at three instances 
during the event of September 30, 2012 (top). The bottom row shows the DEM as a 
function of position and temperature along the line marked in the images at the 
top.}
\label{DEM12}
\end{figure}

The results of our DEM analysis are shown in Figure 
\ref{DEM12} at three instances during the evolution of the event. The 
background plasma now is near $\log T \sim 6.3$ and the cuts shown in the 
bottom row of the Figure cross the loop at two points, at $x=-80$ and $x=-45$; 
the DEM on the loop has a maximum at $\log T\sim 6.86$. Note also the gradual 
cooling of the loop between 07:53~UT and 08:15~UT.

\section{Summary and discussion}
Our analysis of short-lived umbral brightenings from SDO/AIA data showed two cases 
where the brightenings were not confined to the low atmospheric layers. Not 
only were they visible in all AIA channels, which  implies a very extended temperature 
range, but they were also connected by hot coronal loops to opposite-polarity 
footpoints located $\sim50$\arcsec\ away. The association of these 
brightenings with hot coronal loops proves beyond any doubt that their origin is 
magnetic. In this respect, they are similar to the umbral flares reported by Tang 
(\cite{1978SoPh...60..119T}), but they are more compact and of shorter duration. 
The brightenings were strongest in the 1600\,\AA\ AIA band, where their intensity 
exceeded that of the quiet Sun and reached the average level of the plage. They 
did not appear in the HMI images, and we estimate their lowest extent to a height 
between 302\,km and 360\,km.

We note briefly that the association with hot coronal loops is not a general 
property of umbral brightenings. We have found several cases that did not 
exhibit such an association (see, {\it e.g.\/} the bright point at $x=-26, 
y=-274$ at 07:42\,UT in movie 2); these were shorter ($\sim 1$\,min) and probably 
similar to the umbral flashes of Beckers \& Tallant (\cite{1969SoPh....7..351B}). 
We are currently preparing a report on them.

The associated hot coronal loops appeared first in the 94\,\AA\ AIA band, where they 
reached maximum 2 to 6 min after the brightening maximum. Subsequently, 
they became visible in the 335\,\AA\ band and, in the case of the September 30 
2012 event, in the 211\,\AA\ band. In the GOES-15 SXI soft X-ray images the loop 
was visible in all bands with a peak near that of the 94\,\AA\ loop. This 
behavior is suggestive of gradual cooling of hot plasma.

The high temperature of the loops is confirmed not only by their appearance in the 
hot 94\,\AA\ channel and in the SXI images, but also by our DEM analysis, 
which showed that the average temperature of the loop plasma was $\sim 6.5\,10^6$\,K. 
Thus the events presented here are clear examples of plasma cooling from high 
temperatures of several MK to $\approx$2.5\,MK. The fact that the hot 
loops are rooted in sunspot umbrae is a key advantage of our data, since this 
allows direct observation of their footpoints without obscuration from low-lying 
structures. Obscuration is a limiting factor in observations of loops rooted in 
plages, where it is often hard to obtain a view of their footpoints and is difficult 
even to identify them. This is the case for se\-ve\-ral studies of hot evolving 
coronal loops in active regions using Hinode and AIA ({\it e.g.\/} Warren et al. 
\cite{2011ApJ...734...90W}).

Cooling of individual coronal loops or even  of large 2D fields in ARs has been 
observed in the past ({\it e.g.\/} \cite{2003ApJ...593.1174W}; 
\cite{2007ApJ...657.1127L}; \cite{2009ApJ...695..642U}; \cite{2011ApJ...738...24V}; 
Warren et al. \cite{2011ApJ...734...90W}).
There are two possibilities to explain the cooling. One is that we may have a 
quasi-steady hot coronal loop that begins to cool down once its (quasi-steady) 
heating is shut off. This possibility can be safely excluded, because 
quasi-steady multi-million degrees hot coronal loops are expected to have intense footpoint 
emissions ({\it e.g.\/} Patsourakos \& Klimchuk \cite{2008ApJ...689.1406P}). 
However, our observations of the umbral loop footpoints show that they are only 
a factor $\approx$ 2-3 more intense than the associated coronal sections. A 
second, more plausible scenario for the observed evolution is that the coronal-%
loop cooling follows its impulsive heating to multi-million K. It is possible 
that the heating phase cannot be observed due to the low emission measure 
associated with it: the temperature increase is not initially tracked by the loop 
density, because it takes some time to fill a coronal loop with hot plasma via 
chromospheric evaporated material ({\it e.g.\/} Patsourakos \& Klimchuk 
\cite{2006ApJ...647.1452P}).

We point out here that high-frequency impulsive heating, with  a 
repeat time between successive heating pulses shorter than the coronal cooling 
time, also leads to quasi-steady conditions. If this had been the case, one would 
expect re-heating of the observed loops. However, inspection of  AIA movies 
spanning over 90 minutes after the appearance of the observed loops did not show 
strong evidence for this. The fact that the observed loops did not appear to cool 
down to the 211 channel and below may imply that they did not have enough 
emission to stand above the strong, highly-structured and dynamic active-region 
background observed in warmer emissions ($\le2$ MK). Our inferences above would 
be more firmly confirmed or rejected through detailed modeling of the 
observed loops, which is an important task for the future. 

The relatively simple geometry of the hot loops and their footpoints as well as the 
possibility of following their temperature-density evolution through computating 
the DEM from AIA data makes these structures ideal for studying the  heating and 
cooling of loops; we intend to do this in a future work.

\begin{acknowledgements}
The authors made extensive use of the data bases of SDO (AIA and HMI), GONG, GOES,  
CALLISTO, WIND/WAVES, the USAF RSTN network and the {\it Helioviewer} site 
(http://helioviewer.org); they are grateful to all those that worked for the 
development and operation of these instruments and for making the data available. 
The research of SP has been supported in part by the European Union 
(European Social Fund ESF) and in part by the Greek Operational Program 
``Education and Lifelong Learning`` of the National Strategic Reference Framework 
(NSRF) - Research Funding Program: Thales ``Hellenic National Network for Space 
Weather Research``-MIS 377274. S.P. also acknowledges support from an FP7 Marie 
Curie Grant (FP7-PEOPLE-2010-RG/268288).
\end{acknowledgements}

\end{document}